\documentclass[12pt]{article}

\usepackage{amsmath,amssymb,cite}
\usepackage{epsfig,lscape,rotating}
\usepackage{graphicx}
\usepackage{color}

\include{newcommands}

\newcommand{\dmrm}{\mathrm{d}}

\def\beq{\begin{equation}}
\def\eeq#1{\label{#1}\end{equation}}
\def\eeqn{\end{equation}}
\def\beqa{\begin{eqnarray}}
\def\eeqa#1{\label{#1}\end{eqnarray}}
\def\eeqan{\end{eqnarray}}
\def\CR{\nonumber \\ }
\def\leqn#1{(\ref{#1})}

\def\beqs{\begin{equation} \begin{split}}
\def\eeqs#1{\label{#1} \end{split} \end{equation}}
\def\eeqsn{\end{split} \end{equation}}

\def\to{\rightarrow}

\def\to{\rightarrow}

\def\smax{s^{\rm max}}
\newcommand{\bspace}{\!\!\!\!}
\def\met{\mbox{$E{\bspace}/_{T}$}}
\begin{document}

\begin{titlepage}
%\begin{flushright}
%{\tt hep-ph/yymmnnn} \\
%\end{flushright}

\vskip.5cm
\begin{center}
{\huge \bf Testing Gluino Spin with Three-Body Decays}
\vskip.2cm
\end{center}

\begin{center}
{\bf Csaba Cs\'aki, Johannes Heinonen and Maxim Perelstein } \\
\end{center}
\vskip 8pt

\begin{center}

\vspace*{0.2cm}

{\it Institute for High Energy Phenomenology\\
Newman Laboratory of Elementary Particle Physics, \\
Cornell University, Ithaca, NY 14853, USA } \\

\vspace*{0.3cm} {\tt
csaki,heinonen,maxim@lepp.cornell.edu }
\end{center}

\vglue 0.3truecm

\begin{abstract}\vskip 3pt \noindent
We examine the possibility of distinguishing a supersymmetric gluino
from a Kaluza-Klein gluon of universal extra dimensions (UED) at the
Large Hadron Collider (LHC). We focus on the case when all kinematically
allowed tree-level decays of this particle are 3-body decays into two jets
and a massive daughter (typically weak gaugino or Kaluza-Klein weak gauge
boson). We show that the shapes of the dijet invariant mass distributions
differ significantly in the two models, as long as the mass of the decaying
particle $m_A$ is substantially larger than the mass of the massive
daughter $m_B$. We present a simple analysis estimating the number of events
needed to distinguish between the two models under idealized conditions.
For example, for $m_A/m_B=10$, we find the required number of events to be
of order several thousand, which should be available at the LHC within a few
years. This conclusion is confirmed by a parton level Monte Carlo study
which includes the effects of experimental cuts and the combinatoric
background.
\end{abstract}

\end{titlepage}

%%%%%%%%%%%%%%%%%%%%%%%%%%%%%%%%%%%%%%%%%%%%%%%%%%%%%%%%%%%%%%%%%%%%%%%%%%%%%%%%%%%%%%%%%%%
\section{Introduction}
\setcounter{equation}{0}

Very soon, experiments at the Large Hadron Collider (LHC) will begin direct
exploration of physics at the TeV scale. Strong theoretical arguments suggest
that this physics will include new particles and forces not present in the
Standard Model (SM). Several theoretically motivated extensions of the
Standard Model at the TeV scale have been proposed. After new physics
discovery at the LHC, the main task of the experiments will be to determine
which of the proposed models, if any, is correct.

Unfortunately, there exists a broad and well-motivated class of SM extensions
for which this task would be highly non-trivial. In these models, the new
TeV-scale particles carry a new conserved quantum number, not carried by the
SM states. The lightest of the new particles is therefore stable. Furthermore,
the stable particle interacts weakly, providing a very attractive ``weakly
interacting massive particle'' (WIMP) candidate
for dark matter with relic abundance naturally in the observed range.
Models of this class include the minimal supersymmetric standard model
(MSSM) and a variety of other supersymmetric models with conserved R parity,
Little Higgs models with T parity (LHT), and models with universal extra
dimensions
(UED) with Kaluza-Klein (KK) parity. All these models  have a common
signature at a hadron collider: pair-production of new states is followed by
their
prompt decay into visible SM states and the lightest new particle, which
escapes the detector without interactions leading to a ``missing transverse
energy'' signature. If this universal signature is observed at the LHC,
how does one determine which of these models is realized?

One crucial difference between the MSSM and models such as LHT or UED is the
correlation between spins of the new particles and their gauge charges.
In all these models, all (or many of) the new states at the TeV scale
can be paired up with the known SM particles, with particles in the same
pair carrying identical gauge charges. However, while in the LHT and UED
models the two members of the pair have the same spin, in the MSSM and other
supersymmetric models their spins differ by 1/2. Thus, measuring the
spin of the observed new particles provides a way to discriminate among
models.

Experimental determination of the spin of a heavy unstable particle with
one or more invisible daughter(s) in hadron collider environment
is a difficult task. One possibile approach, which
recently received considerable attention in the
literature~\cite{cascade1,cascade2,cascade3,Athanasiou:2006ef,WY1,WY2,tilman},
is to use angular correlations between the observable particles emitted in
subsequent steps of a cascade decay, which are sensitive to intermediate
particle spins. This strategy is promising, but its success depends on the
availability of long cascade decay chains, which may or may not occur
depending on the details of the new physics spectrum. It is worth
thinking about other possible strategies for spin determination.

In this paper, we explore the possibility of using 3-body decays of heavy
new particles to determine their spin. The most interesting example is the
3-body decay of the MSSM gluino into a quark-antiquark pair and a weak
gaugino,
\beq
\tilde{g}\to q+\bar{q}+\chi.
\eeq{process}
In a large part of the MSSM parameter space,
this decay has a large branching ratio: this occurs whenever all squarks
are heavier than the gluino.\footnote{The main competing gluino decay
channel in this parameter region is a two-body decay $\tilde{g}\to g\chi$,
which first arises at one-loop level and generically has a partial width
comparable to the tree-level decay~\leqn{process}. The gluino decay patterns
in this parameter region have been analyzed in detail in Ref.~\cite{TW}.}
Under the same condition, gluino pair-production dominates SUSY signal at
the LHC. We will argue that the invariant mass distribution of the jets
produced in reaction~\leqn{process} contains non-trivial information about
the gluino spin, and can be used to distinguish this process from, for
example, its UED counterpart, $g^1\to q+\bar{q}+B^1/W^1$.

It is important to note that the jet invariant mass distribution we study
depends not just on the spin of the decaying particle, but also on the
helicity structure of the couplings which appear in the decay~\leqn{process},
as well as on the masses of the decaying particle, the invisible daughter,
and the off-shell particles mediating the decay. If all these parameters were
measured independently, the jet invariant mass distribution would
unambiguously determine the spin. However, independent determination of many
of the relevant parameters will be very difficult or impossible at the LHC.
In this situation, proving the spin-1/2 nature of the decaying particle
requires demonstrating that the experimentally observed curve cannot be
fitted with any of the curves predicted by models with other spin assignments,
independently of the values of the unknown parameters. This considerably
complicates our task. Still, interesting information can be extraced.
For example, we will show that, even if complete ignorance of the decaying
and intermediate particle masses is assumed, the jet invariant mass
distribution allows one to distinguish between the decay~\leqn{process} in the
MSSM and its UED counterpart (assuming the couplings specified by each model)
at the LHC.

The paper is organized as follows. After setting up our notation and reviewing
the basics of three-body kinematics in Section 2, we present a simple toy model
showing how dijet invariant mass distributions from three-body decays can
be used to probe the nature of the decaying particle and its couplings in
Section 3. Section 4 discusses using this observable for MSSM/UED
discrimination, and contains the main results of the paper. Section 5 contains
the conclusions. Appendix A contains the polarization analysis of the decay
$g^1\to q+\bar{q}+B^1/W^1$ in UED, which sheds some light on the main 
features of the dijet invariant mass distribution in this case. Appendix B
contains a brief review of the Kullback-Leibler distance, a statistical 
measure used in our analysis. 

\section{The Setup and Kinematics}
\setcounter{equation}{0}

We are interested in three-body decays of the type
\beq
  A \rightarrow q + \bar{q} + B,
\eeq{decay}
where $A$ and $B$ are TeV-scale particles. The main focus of this paper will
be on the case when $A$ is the gluino of the MSSM or the KK gluon of the UED
model, and $B$ is a neutalino or chargino of the MSSM or a KK electroweak
gauge boson of the UED; however the discussion in this section applies more
generally. We assume that $q$ and $\bar{q}$ are massless, and denote their
four-momenta by $p_1$ and $p_2$ respectively. To describe the kinematics in
Lorentz-invariant terms, we introduce the ``Mandelstam variables'',
\beqa
s &=& (p_1+p_2)^2 = (p_A-p_B)^2 \,,\CR
u &=& (p_1+p_B)^2 = (p_A-p_2)^2 \,,\CR
t &=& (p_2+p_B)^2 = (p_A-p_1)^2 \,,
\eeqa{Man}
of which only two are independent since
\beq
s+t+u=m_A^2 + m_B^2.
\eeq{Man_rel}
The allowed ranges for the Mandelstam variables are determined by energy and
momentum conservation; in particular,
\beq
0\leq s \leq \smax \equiv (m_A-m_B)^2.
\eeq{smax}
We will assume that $p_B$ cannot be reconstructed, either because $B$ is
unobservable or is unstable with all decays containing unobservable daughters.
Moreover, since the parton center-of-mass frame is unknown, no information is
available about the motion of particle $A$ in the lab frame. Due to these
limitations, the analysis should use observables that can be reconstructed
purely by measuring the jet four-momenta, and are independent of the velocity
of $A$ in the lab frame. The only such observable is $s$, and the object of
interest to us is the distribution $d\Gamma/ds$. This is given by
\beq
\frac{d\Gamma}{ds}\,=\,\frac{1}{64\pi^3}\frac{s}{m_A^2}\,
\int_{E_B-p_B}^{E_B+p_B} \frac{dy}{(m_A-y)^2}\,\bar{|{\cal M}|^2}\,,
\eeq{master}
where
\beqa
E_B &=& \frac{m_A^2+m_B^2-s}{2m_A}\,,\CR
p_B &=& \sqrt{E_B^2-m_B^2}\,,
\eeqa{e3p3}
and ${\cal M}$ is the invariant matrix element for the decay~\leqn{decay},
with the bar denoting the usual summation over the final state spins and
other quantum numbers and averaging over the polarization and other
quantum numbers of $A$.\footnote{This procedure should take into account the
polarization of $A$ if it is produced in a polarized state. In the examples of
this paper, production is dominated by strong interactions and $A$ will
always be produced unpolarized.} The quantity $\bar{|{\cal M}|^2}$ can be
expressed in terms of the variables~\leqn{Man}; substitutions
\beq
t \to m_A^2 -\frac{sm_A}{m_A-y},~~~
u\to \frac{sy}{m_A-y}+m_B^2
\eeq{subs}
should be made in $\bar{|{\cal M}|^2}$ before performing the
integral in Eq.~\leqn{master}. Notice that Eq.~\leqn{master} is valid in the
rest frame of the particle $A$; however, since $s$ is Lorentz-invariant, its
Lorentz transformation is a trivial overall rescaling by time dilation, and
the shape of the distribution is unaffected. The strategy we will pursue is to
use this shape to extract information about the decay matrix element
${\cal M}$, which is in turn determined by the spins and couplings of the
particles $A$ and $B$.

To separate the effects of non-trivial structure of the decay matrix element
>from those due merely to kinematics, it will be useful to compare the dijet
invariant mass distributions predicted by various theories to the ``pure phase
space'' distribution, obtained by setting the matrix element to a constant
value. From Eq.~\leqn{master}, the phase space distribution is given by
\beq
\frac{d\Gamma}{ds}\,=\,\frac{1}{32\pi^3}\,\frac{|{\bf p}_B|}{m_A} \propto
\sqrt{(s-m_A^2-m_B^2)^2-4m_A^2 m_B^2}.
\eeq{phasespace}
This distribution\footnote{Since we are concerned with the shapes
of the dijet invariant mass distributions in various models and not their
overall normalizations, all distributions appearing on the plots
throughout this paper are
normalized to have the same partial width $\Gamma=\int_0^{\smax}
\frac{d\Gamma}{ds}\,ds$.} is shown by a solid black line in
Fig.~\ref{fig:toy}. Notice that the phase space distribution has an
endpoint at $s=\smax$, with the asymptotic behavior given by
\beq
\frac{d\Gamma}{ds} \sim (s-\smax)^{1/2}
\eeq{PSendpoint}
as the endpoint is approached.

\section{Chiral Structure in Three-Body Decays: a Toy Model \label{sec:toy}}
\setcounter{equation}{0}

\begin{figure}
\begin{center}
\includegraphics[width=6cm,keepaspectratio=true]{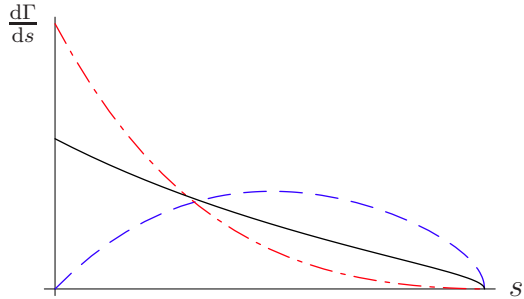}
\put(5,0){$s$}
\put(-185,100){$\frac{\dmrm \Gamma}{\dmrm s}$}
\caption{Dijet invariant mass distribution for the toy model 1 (blue/dashed)
and model 2 (red/dot-dashed) compared to phase space (black/solid) for
$M_*/m_A=1.5$ and $m_B/m_A=0.1$.\label{fig:toy} }
\end{center}
\end{figure}

\begin{figure}
\centering
\includegraphics[width=15cm,keepaspectratio=true]{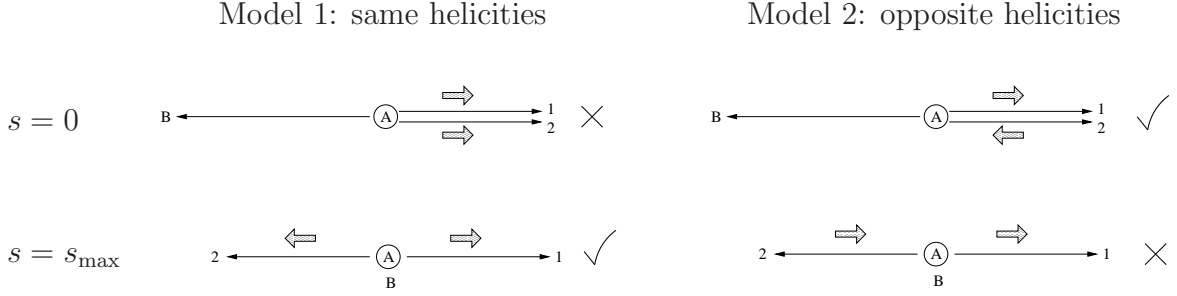}
\put(-440,60){$s = 0$}
\put(-440,10){$s = s_\text{max}$}
\put(-360,100){Model 1: same helicities}
\put(-160,100){Model 2: opposite helicities}
\caption{Momenta (long arrows) and helicities (short arrows) in the $A$ rest
frame for $s=0$ and $s=\smax$ in the two toy models of section 3.
\label{fig:AM}}
\end{figure}

To illustrate how the chiral structure of the couplings involved
in the decay~\leqn{decay} can be determined from the dijet invariant mass
distribution, consider a situation when the particles $A$ and $B$ are real
scalars. Introduce a massive Dirac fermion $\Psi$ of mass $M_*>m_A$, and
consider the following two models: model 1 defined by
\beq
{\cal L}_1\,=\, y_A A \bar{\Psi} P_L q + y_B B
\bar{\Psi} P_R q + {\rm h.c.}
\eeq{toy1}
and model 2 defined by
\beq
{\cal L}_2\,=\, y_A A \bar{\Psi} P_L q + y_B B
\bar{\Psi} P_L q + {\rm h.c.}
\eeq{toy2}
The matrix element for the decay~\leqn{decay} in model 1 is given by
\beq
\sum_{\rm spin}|{\cal M}_1|^2\,=\, 2 y_A^2 y_B^2 (M_*^2\, s) \,
\left( \frac{1}{(t-M_*^2)^2}+\frac{1}{(u-M_*^2)^2} \right)\,,
\eeq{m1}
while in model 2 it is given by
\beq
\sum_{\rm spin}|{\cal M}_2|^2\,=\, 2 y_A^2 y_B^2 \left(
(m_A^2+m_B^2) t u -m_A^2 m_B^2 \right) \,
\left( \frac{1}{t-M_*^2}+\frac{1}{u-M_*^2} \right)^2\,.
\eeq{m2}
The dijet invariant mass distributions in the two models are shown
by the blue/dashed line (model 1) and red/dot-dashed line (model 2) in
Fig.~\ref{fig:toy}. Their strikingly different shapes are due to the
angular momentum conservation law and to the different helicity structure of
the couplings. To understand this, consider this decay
in the $A$ rest frame. In this frame, $s=2E_1E_2(1-\cos\theta_{12}).$ When
$s=0$, the quark and the antiquark travel in the same direction, as
illustrated in Fig.~\ref{fig:AM}. Since $A$ and $B$ have zero spin, the sum of
the quark and antiquark helicities must vanish for this kinematics.
In model 1, the quark and the antiquark have the same helicity, and the
decay is forbidden for $s=0$; in model 2, it is allowed. In contrast, when
$s=\smax$, the particle $B$ is at rest, and the quark and the antiquark
travel in the opposite directions. By angular momentum conservation, their
helicities must be equal. In model 1, this is the case, and the distribution
approaches that of pure phase space in the limit $s\to\smax$. In model 2, this
kinematics is forbidden, the matrix element vanishes at the endpoint, and
the distribution behaves as $d\Gamma/ds \propto (s-\smax)^{3/2}$.

\section{Model Discrimination: SUSY Versus UED}
\setcounter{equation}{0}

In this Section, we will show that measuring the shape of the dijet invariant
mass distribution arising from a three-body decay of a heavy colored
particle may allow to determine whether the decaying particle is the gluino
of the MSSM or the KK gluon of the UED model. We will begin by comparing
the analytic predictions for the shapes of the two distributions at leading
order. We will then present a parton-level Monte Carlo study which demonstrates
that the discriminating power of this analysis persists after the main
experimental complications (such as the combinatioric background, finite
energy resolution of the detector, and cuts imposed to suppress SM
backgrounds) are taken into account.

%%%%%%%%%%%%%%%%%%%%%%%%%%%%%%%%%%%%%%%%%%%%%%%%%%%
\subsection{Gluino decay in the MSSM}

\begin{figure}
\centering
\includegraphics[width=12cm,keepaspectratio=true]{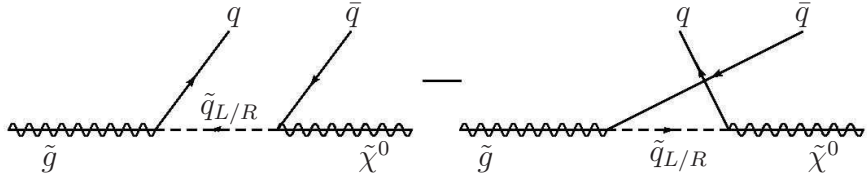}
  % for 1st diagram
  \put(-250,55){$q$}
  \put(-205,55){$\bar{q}$}
  \put(-320,0){$\tilde{g}$}
  \put(-200,0){$\tilde{\chi}^0$}
  \put(-260,20){$\tilde{q}_{L/R}$}
  % for 2nd diagram
   \put(-80,55){$q$}
  \put(-35,55){$\bar{q}$}
  \put(-155,0){$\tilde{g}$}
  \put(-30,0){$\tilde{\chi}^0$}
  \put(-90,2){$\tilde{q}_{L/R}$}
  \caption{The Feynman diagrams for gluino three-body decay in the MSSM.
\label{fig:MSSM_fd}}
\end{figure}

We consider the MSSM in the region of the parameter space where all squarks
are heavier than the gluino, forbidding the two-body decays $\tilde{g}\to
\tilde{q} q$. In this situation, gluino decays through three-body channels.
We study the channel
\beq
\tilde{g}(p_A) \to q(p_1)+\bar{q}(p_2)+\tilde{\chi}_1^0(p_B),
\eeq{gluino_dec}
where $q$ and $\bar{q}$ are light (1st and 2nd generation) quarks, and
$\tilde{\chi}_1^0$ is the lightest neutralino which we assume to be
the LSP. (Note that many of our results would continue to hold if
 $\tilde{\chi}_1^0$ is replaced with a heavier neutralino or a chargino.
The only extra complication in these cases would be a possible additional
contribution to the combinatoric background from the subsequent cascade
decay of these particles.) The leading-order Feynman diagrams for the
process~\leqn{gluino_dec} are shown in Fig.~\ref{fig:MSSM_fd}; the vertices
entering these diagrams are well known (see for example Ref.~\cite{HK}).
The spin-summed and averaged matrix element-squared has the form (up to
an overall normalization constant)
\beq
\sum_{\rm spin}|{\cal M}_{\rm MSSM}|^2\,=\, |C_L|^2 F(s,t,u;M_{L*})   +
|C_R|^2 F(s,t,u;M_{R*})\,,
\eeq{MSSM_M}
where
\beq
  F(s,t,u; M) =  \frac{(m_A^2-t)(t-m_B^2)}{(t-M^2)^2} +
\frac{(m_A^2-u)(u-m_B^2)}{(u-M^2)^2} + 2 \frac{m_A m_B s}{(u-M^2)(t-M^2)}\,.
\eeq{MSSM_F}
Here $m_A$, $m_B$, $M_{L*}$ and $M_{R*}$ are the masses of the gluino, the
neutralino, the squarks $\tilde{q}_L$ and $\tilde{q}_R$, respectively.
In order to keep the analysis general, we will not assume any relationships
(such as mSUGRA contraints) among these parameters, and will always work in
terms of weak-scale masses. We also define
\beqa
C_L & = & T^3_q N_{12} - t_w (T^3_q - Q_q) N_{11}\,, \CR
C_R & = & t_w Q_q N_{11}\,,
\eeqa{CLandCR}
where $T^3_u=+1/2, T^3_d=-1/2, Q_u=+2/3, Q_d=-1/3, t_w=\tan\theta_w$, and
$N$ is the neutralino mixing matrix\footnote{We assume that $N$ is real.
It is always possible to redefine the neutralino fields to achieve this.
However one should keep in mind that the neutralino eigenmasses may be
negative with this choice.} in the basis $(\tilde{B}, \tilde{W}^3,
\tilde{H}_u^0, \tilde{H}_d^0)$. We have neglected the mixing
between the left-handed and right-handed squarks, which is expected to be
small in the MSSM.\footnote{Large mixing in the stop sector may be present, and
is actually preferred by fine-tuning arguments in the MSSM (see, e.g.,
Ref.~\cite{PS}). However events with
top quarks in the final state are characterized by more complicated
topologies and can be experimentally distinguished from the events with light
quarks that we are focussing on here.} Since up and down type quarks are
experimentally indistinguishable, the dijet invariant mass distribution
$d\Gamma/ds$ should include both the contributions of up-type and
down-type squarks.

%%%%%%%%%%%%%%%%%%%%%%%%%%%%%%%%%%%%%%%%%%%%%%%%%%%
\subsection{Decay of the gluon KK mode in the UED model}
\label{UED calculation}

\begin{figure}
 \centering
   \includegraphics[width=12cm,keepaspectratio=true]{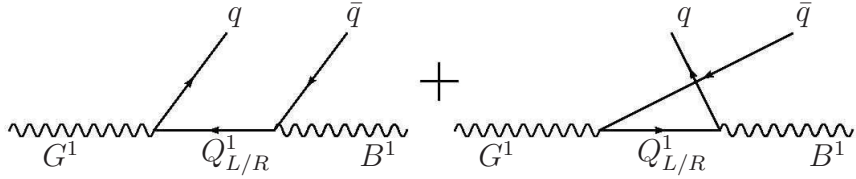}
  % for 1st diagram
  \put(-250,55){$q$}
  \put(-205,55){$\bar{q}$}
  \put(-320,0){$G^1$}
  \put(-200,0){$B^1$}
  \put(-260,2){$Q_{L/R}^1$}
  % for 2nd diagram
   \put(-80,55){$q$}
  \put(-35,55){$\bar{q}$}
  \put(-155,0){$G^1$}
  \put(-30,0){$B^1$}
  \put(-95,2){$Q_{L/R}^1$}
  \caption{The Feynman diagrams for the KK gluon three-body decay in UED.
\label{fig:UED_fd}}
\end{figure}

The counterpart of the decay~\leqn{gluino_dec} in the universal extra
dimensions (UED) model is the decay
\beq
g^1(p_A) \to q(p_1)+\bar{q}(p_2)+B^1(p_B),
\eeq{KKglue_dec}
where $g^1$ and $B^1$ are the first-level Kaluza-Klein (KK) excitations of the
gluon and the hypercharge gauge boson, respectively. We ignore the mixing
between $B^1$ and the KK mode of the $W^3$ field, which is small provided that
the radius of the extra dimension is small, $R\ll 1/M_W$, and assume that
the $B^1$ is the LTP. As in the MSSM
case, the decay~\leqn{KKglue_dec} is expected to have a substantial branching
fraction when all KK quarks $Q^1_R$ and $Q^1_L$ are heavier than the KK
gluon. Note that in the original UED model~\cite{ued1}, the KK modes of all
SM states were predicted to be closely degenerate in mass around $M=1/R$; it
was however later understood~\cite{ued2} that kinetic terms localized on the
boundaries of
the extra dimension can produce large mass splittings in the KK spectrum.
Since such kinetic terms are consistent with all symmetries of the theory,
we will assume that they are indeed present, and treat the masses of the
$g^1$, $B^1$, $Q^1_R$ and $Q^1_L$ fields as free parameters.

The leading-order Feynman diagrams for the decay~\leqn{KKglue_dec} are
shown in Fig.~\ref{fig:UED_fd}. (We ignored the contribution of the 
diagrams mediated by $Q_{L/R}^i$ with $i\geq 2$, which are suppressed by the
larger masses of the higher KK modes.) The relevant couplings have the form
\beqa
 & & g_3 G^1_\mu \left[ \bar{q} \gamma^\mu P_R Q^1_R + \bar{q}
\gamma^\mu P_L Q^1_L + \bar{Q}^1_R \gamma^\mu P_R q +
\bar{Q}^1_L \gamma^\mu P_L q  \right] \,+\CR
& & g_1 B^1_\mu \left[ Y(q_R)\,\bar{q} \gamma^\mu P_R Q^1_R + Y(q_L)\,
\bar{q} \gamma^\mu P_L Q^1_L + Y(q_R) \,\bar{Q}^1_R \gamma^\mu P_R q +
Y(q_L) \,\bar{Q}^1_L \gamma^\mu P_L q  \right],
\eeqa{UED_fr}
where $Y(q_L)=1/6, Y(u_R)=+2/3$ and $Y(d_R)=-1/3$ are the
hypercharges.\footnote{The structure of the couplings between the
KK gauge bosons and SM
(or KK) quarks are unaffected by brane-localized kinetic terms as long as
these terms are flavor-independent.}
The spin-summed and averaged matrix element-squared has the form (up to
an overall normalization constant)
\beq
\sum_{\rm spin}|{\cal M}_{\rm UED}|^2\,=\, Y_L^2 \,G(s,t,u;M_{L*}) +
Y_R^2 \,G(s,t,u;M_{R*})\, ,
\eeq{UED_M}
where $M_{L*}$ and $M_{R*}$ are the masses of the left- and right-handed
quark KK modes $Q^1_L$ and $Q^1_R$, and
\beq
G(s,t,u;M)=\frac{h_1(s,t,u)}{(t-M^2)^2} + \frac{h_1(s,u,t)}{(u-M^2)^2} + 2 \frac{h_2(s,t,u)}{(t-M^2)(u-M^2)},
\eeq{UED_G}
with
\beqa
  h_1(s,t,u) &=& 4(t u - m_A^2 m_B^2) + \frac{t^2}{m_A^2 m_B^2}
\left( 2 s (m_A^2+m_B^2) + t u -m_A^2 m_B^2 \right)\,,\CR
  h_2(s,t,u) &=& 4s(m_A^2+m_B^2) - \frac{t u}{m_A^2 m_B^2} \left(
2 s (m_A^2+m_B^2)+t u-m_A^2 m_B^2  \right)\,.
\eeqa{UED_h}

\subsection{Model Discrimination: a Simplified Analysis}
\label{simple}

\begin{figure}
\begin{center}
\includegraphics[width=6cm,keepaspectratio=true]{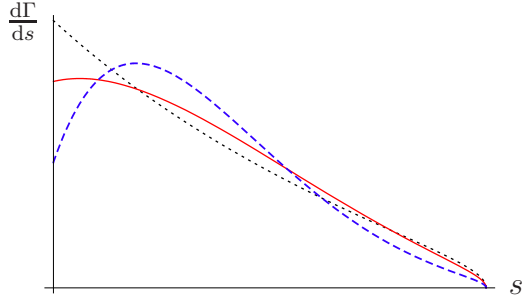}
\put(5,0){$s$}
\put(-185,100){$\frac{\dmrm \Gamma}{\dmrm s}$}
\caption{Dijet invariant mass distribution for the UED (blue/dashed)
and the MSSM (red/solid) models, compared to pure phase space (black/dotted)
for $M_{L*}/m_A=M_{R*}/m_A=1.5$ and $m_B/m_A=0.1$.\label{fig:dGdsall}}
\end{center}

\end{figure}

Armed with the expressions~\leqn{MSSM_M} and~\leqn{UED_M}, it is
straightforward to obtain the dijet invariant mass distributions for
gluino and KK gluon decays and compare them. For example, the two
distributions for a particular choice of parameters, along with the
pure phase space distribution, are shown in Fig.~\ref{fig:dGdsall}.
While not as strikingly different as the two toy models of
Section~\ref{sec:toy}, the curves predicted by the MSSM and the UED
are clearly distinct. (The suppression of the UED distribution
compared to phase space around $s=0$ and $s=s_{max}$ can be easily
understood using angular momentum conservation, as explained
in Appendix~\ref{app:pol}.) In this section, we will
perform a simplified analysis of the discriminating power of these
distributions, ignoring experimental complications such as cuts,
finite energy resolution, combinatoric and SM backgrounds, and
systematic errors. We will return to include some of these
complications in the following section.

The distrubution in each model depends on a number
of parameters, including the mass of the mother particle $m_A$, the mass of
the invisible daughter $m_B$, and the masses of intermediate particles:
$(\tilde{u}_L, \tilde{d}_L, \tilde{u}_R, \tilde{d}_R)$ in the MSSM case and
$(U^1_L, D^1_L, U^1_R, D^1_R)$ in the UED case. We assume that the
partners of the up-type quarks of the first two generations and the
down-type quarks for all three generations are degenerate, and do not 
include the diagrams with intermediate stops (or KK tops) since they produce
tops in the final state. Furthermore, since the Yukawa couplings
for the first two generations are small, it is safe to assume that
$m(\tilde{u}_L)=m(\tilde{d}_L)$ in the MSSM and $m(U^1_L)=m(D^1_L)$ in UED.
Since an overall rescaling of all masses does not affect the shape of the
distribution, we need four dimensionless parameters to specify the
mass spectrum in each model; we use the particle masses in units of $m_A$.
Experimentally, these four parameters may be very difficult to obtain
independently. A direct measurement of the masses of squarks/KK quarks
%that appear as intermediate states in the process we study
may well be impossible, since these
particles may be too heavy to be produced on-shell. Also,
while it is easy to measure $m_A-m_B$ (one can use the endpoint of the
dijet invariant mass distribution or other simple observables such as the
effective mass~\cite{Meff} or its variations~\cite{Mt2}),
it is much more difficult to measure $m_A$ and $m_B$ individually~\cite{MR},
which would be required in order to obtain $m_B/m_A$. In this study, we will
conservatively assume no prior knowledge of any of these parameters. (Of
course, if some independent information about them is available,
for example the overall mass scale is constrained by production cross section
considerations, this information can be folded into our analysis, increasing
its discriminating power.) In addition to the unknown masses, the matrix
elements in the MSSM depend on the neutralino mixing matrix elements, $N_{11}$
and $N_{12}$, although only the ratio $N_{11}/N_{12}$ affects the shape of the
distribution. Again,
this parameter is difficult to measure at the LHC, and we will assume that it
is unknown; fortunately, the effect of varying it is quite small.

To quantify the discriminating power of the proposed observable, we use
the following procedure. We assume that
the experimental data is described by the MSSM curve with a
particular set of parameters.\footnote{We checked that the results of our
analysis are approximately independent of which model, MSSM or UED, is
assumed to be the ``true'' one.} We then ask, how many events (assuming
statistical errors only) would be
required to rule out the UED as an explanation of this distribution? To
answer this question, we scan over 50000 points in the UED parameter space:
\beqa
m_B/m_A &=& (0\ldots 0.5),~~M(Q_L^1)/m_A=(1.05\ldots 3.0),\CR
M(D_R^1)/m_A &=&(1.05\ldots 3.0),~~M(U_R^1)/m_A=(1.05\ldots 3.0).
\eeqa{scan}
For each point in the scan, we compute the Kullback-Leibler (KL) distance (see
Appendix~\ref{app:KL}) between the UED distribution with the parameters at
that point, and the ``experimental'' distribution. We then find the
``best-fit UED'' point, which is the point that gives the smallest KL
distance among the scanned sample. Finally, we compute the number of events
required to rule out the best-fit UED point at a desired confidence level.

\begin{figure}
\centering
\includegraphics[width=1.0cm,keepaspectratio=true]{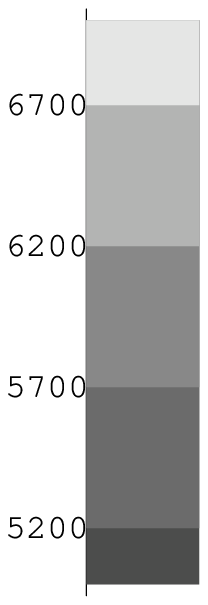}
\hspace{0.5cm}
\includegraphics[width=7cm,keepaspectratio=true]{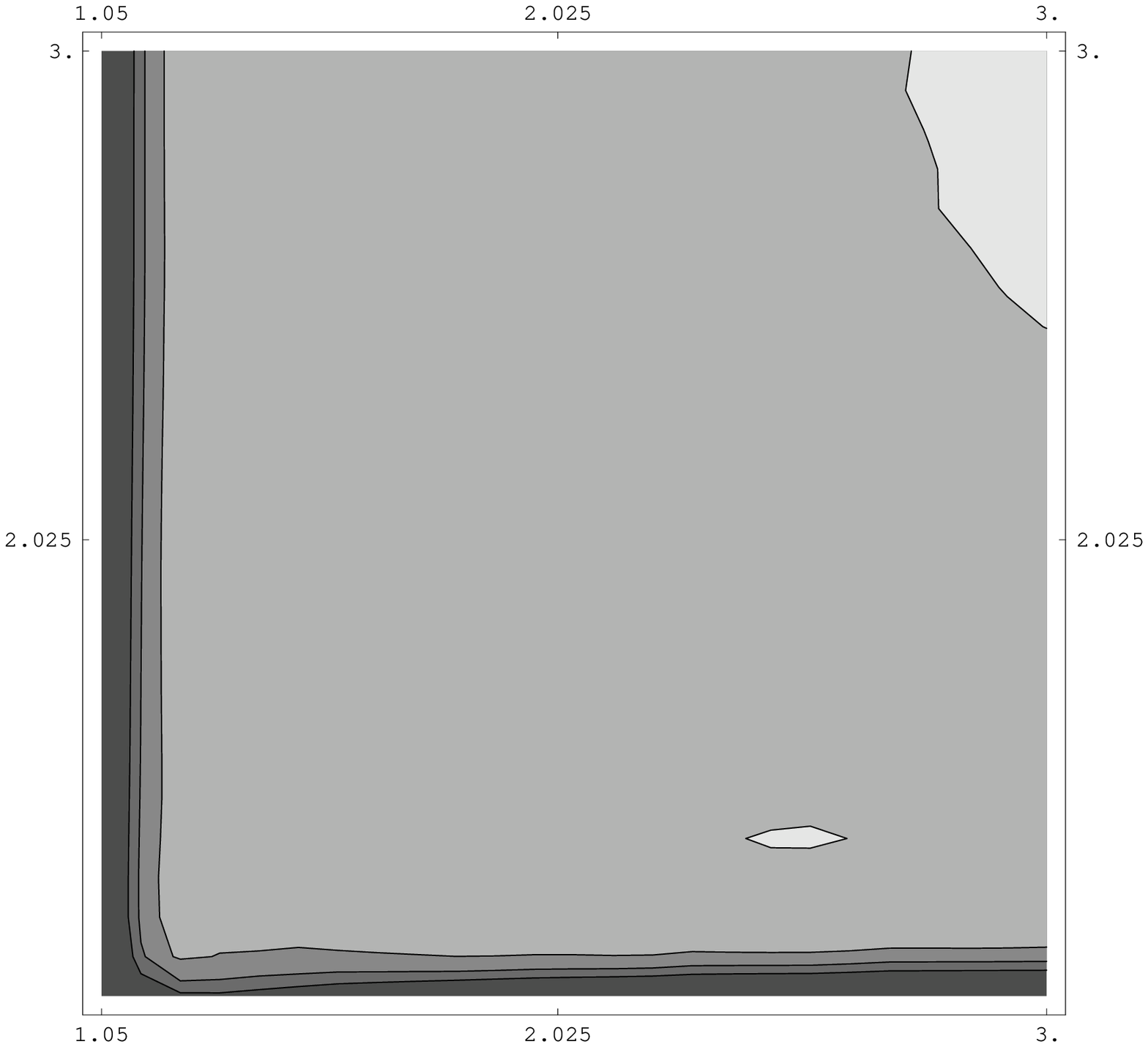}
\caption{Number of events required to distinguish the MSSM and the
UED models based on the invariant mass distributions of jets from
three-body $\tilde{g}/G^1$ decays. \label{fig:scan}}
\end{figure}

The results of this analysis are shown in Fig.~\ref{fig:scan}. The MSSM
parameters used to generate the ``data'' are: $m_B=0.1 m_A$,
$m(\tilde{u}_R)=m(\tilde{d}_R)\equiv m_R, m(\tilde{u}_L)=m(\tilde{d}_L)
\equiv m_L$, $N_{11}/N_{12}=1.$ The parameters $m_L$ and $m_R$ were then
scanned between $1.05m_A$ and $2m_A$, and for each point in the scan the
procedure described in the previous paragraph was performed.
Fig.~\ref{fig:scan} shows the number of events required to rule out the UED
interpretation of the signal at the 99.9\% c.l. (In the language of
Appendix~\ref{app:KL}, this corresponds to $R=1000$.) In a typical point in
the model parameter space, about 6000 events are required. For
comparison, the pair-production cross section for a 1 TeV gluino at the LHC
is about 600 fb, corresponding to 12000 gluinos/year at the initial design
luminosity of 10 fb$^{-1}$/year. The number of events useful for the
measurement studied here depends on the branching ratio of the
decay~\leqn{process}. Since this branching ratio is generically of order one,
we expect $O(10^3)$ useful events/year at the initial stages of the LHC
running. Thus, at least under the highly idealized conditions of this
simplified analysis, this method of model discrimination is quite promising
in a wide range of reasonable model parameters.

\if
The MSSM and UED distributions look more alike when the
squark-gluino mass splitting is small; as a result, more events are
required in that region. As the squarks become heavier, however, the
distributions quickly converge to their asymptotic shapes, and the
discriminating power is approximately independent of the squark masses in most
of the scanned region.
\fi

\begin{figure}
\centering
\includegraphics[width=7cm,keepaspectratio=true]{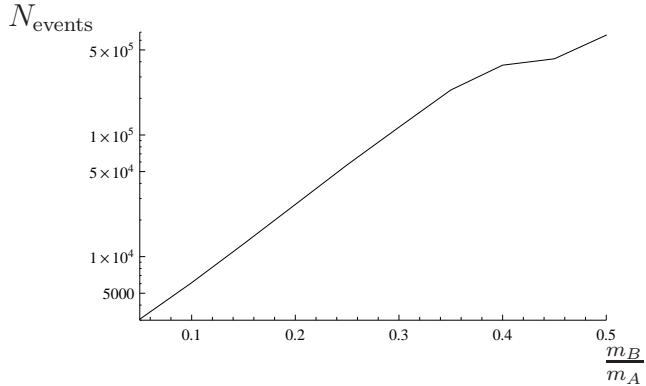}
\put(-5,-10){$\frac{m_B}{m_A}$}
\put(-230,120){$N_\text{events}$}
\caption{Number of events required to distinguish the MSSM and the UED
models, as a function of $m_B/m_A$ of the ``true'' model.}
\label{fig:mBmA}
\end{figure}

\begin{figure}
\centering
\includegraphics[width=7cm,keepaspectratio=true]{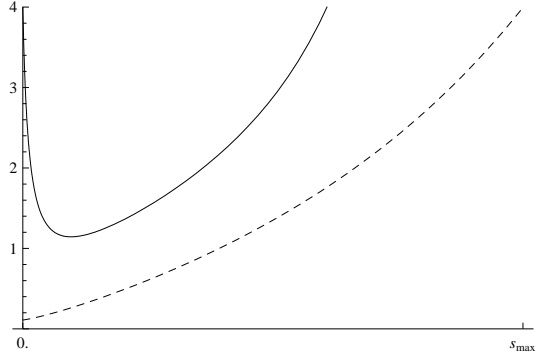}
\caption{Ratio of the decay distributions of $A$ into the
longitudinal component of $B$ to the decay distributions into the
transverse components of $B$ for $m_B/m_A=0.1$(solid) and
$m_B/m_A=0.5$(dashed). For low $m_B/m_A$ the daughter particle is highly
boosted at $s=0$ and will mainly be longitudinally polarized. As
$m_B$ increases, the transeverse polarization becomes more important.}
\label{fig:polratio}
\end{figure}

We checked that the conclusions of this analysis are approximately
independent of the value of $N_{11}/N_{12}$ used to generate the
``data''. They do, however, depend sensitively on the ratio
$m_B/m_A$: as $m_B/m_A$ grows, the MSSM and UED distributions become
more and more alike. This is illustrated in Fig.~\ref{fig:mBmA},
which shows the number of events needed to rule out the ``wrong''
model (assumed to be UED) at the 99.9\% c.l., as a function of
$m_B/m_A$ of the ``true'' model (assumed to be the MSSM with
$m(\tilde{u}_R)=m(\tilde{d}_R) = m(\tilde{u}_L)=m(\tilde{d}_L) =1.5
m_A$ and $N_{11}/N_{12}=1$). The UED scan parameters are the same as
in Eq.~\leqn{scan}, except that we vary $m_B/m_A=(0\ldots 0.9)$ in
this case. It is clear that the discriminating power of the dijet
invariant mass distribution falls rapidly (approximately
exponentially) with growing $m_B/m_A$.  This can be understood as
follows. The main feature of the invariant mass distributions that 
allows for model discrimination is the presence of the sharp dip at 
$s=0$ in the UED case. 
%The Goldstone boson equivalence theorem states that at high
%energies a massive gauge boson becomes equivalent to the Goldstone
%degree of freedom , since the Goldstones couple to momentum.
According to the Goldstone boson equivalence theorem, 
if the daughter particle $B$ in the UED case is highly boosted, the decays 
into its longitudinal component will dominate. The particle $B$ is
highly boosted in the vicintiy of $s=0$, provided that the mass ratio 
$m_B/m_A$ is small; as $m_B/m_A$ grows, the boost becomes less 
pronounced and the decays into the longitudinal component of $B$ are 
less dominant. This is illustrated in Fig.~\ref{fig:polratio}, which 
compares the ratio of partial decay rates into the longitudinal and
transverse modes of $B$ for $m_B/m_A=0.1$ and $m_B/m_A=0.5$. 
However, it is exactly the decays into the longitudinal mode of $B$ that 
are mainly responsible for the characteristic dip at $s=0$; this feature
is far less pronounced for the decays into transverse modes. This means that 
as $m_B/m_A$ is increased, the dip gradually disappears, and the 
discriminating power of our observable fades away.

\if
(so that the $B$ particle is actually highly boosted). In
figure~\ref{fig:polratio} we can see the ratio of the decay
distribution into the longitudinal modes to the decay modes into the
transverse modes of $B$ for two different ratios $m_A/m_B$. For
$m_A/m_B=0.1$ we can clearly see that the longitudinal modes
dominates for low $s$, while for $m_A/m_B=0.5$ the decay into
transverse modes is enhanced. However, it is exactly the decays into
the longitudinal mode that are responsible for the characteristic
dip of the distribution around $s=0$ which lets us distinguish the
UED case from the MSSM case. This means that the characteristic dip
will disappear if we increase $m_A/m_B$, and will our discriminating
power too.
\fi

\subsection{Model Discrimination: a Test-Case Monte Carlo Study
\label{MC}}

Given the large number of simplifying assumptions made in the analysis of
the previous section, a skeptical reader may well wonder how meaningful the
results presented above are. In this section, we will repeat the analysis in a
more realistic setting: effects of experimental cuts and combinatoric
background will be included. We will also bin the distributions, to
approximate the effects of finite jet energy resolution. Since this
analysis involves generating large samples of Monte Carlo (MC) events for each
model, we were not able to perform a scan over the model parameter space,
as we did in the previous section. Instead, we will present a test case,
comparing the MSSM distribution for a single point in the MSSM parameter
space with the distribution generated by the ``best-fit'' UED model for
that point.

The chosen MSSM point has the following parameters: $m_A=1$ TeV, $m_B=0.1
m_A=100$ GeV, $M(\tilde{Q}_L)=M(\tilde{u}_R)=M(\tilde{d}_R)=1.5$ TeV. The
corresponding ``best-fit'' UED point, found by the procedure described in
the previous section, has the following parameters: $m_A=1.06$ TeV,
$m_B=0.15m_A=160$ GeV, $M(Q^1_L)=M(u^1_R)=M(d^1_R)=1.6$ TeV. (Note that the
value of $m_A-m_B$, which can be determined independently, is the same for
these two points.) Using
{\tt MadGraph/MadEvent v4.1}~\cite{MG} event generator, we have simulated a
statistically significant sample (about 20000) of parton-level Monte Carlo
events for each model in $pp$ collisions at $\sqrt{s}=14$ TeV. The simulated
processes are
\beq
pp \to qq\bar{q}\bar{q} \chi^0_1 \chi^0_1
\eeq{MGprocessSUSY}
in the MSSM, and its counterpart,
\beq
pp \to qq\bar{q}\bar{q} B^1 B^1,
\eeq{MGprocessUED}
in UED. With the chosen model parameters, the dominant contribution to the
processes~\leqn{MGprocessSUSY} and~\leqn{MGprocessUED} comes from
pair-production of $\tilde{g}/G^1$,
followed by the three-body decay~\leqn{process}, which is of primary interest
to us. In the MC simulation, we did not demand that the $\tilde{g}/G^1$ be
on-shell; the full tree-level matrix elements for the $2\to 6$
reactions~\leqn{MGprocessSUSY} and~\leqn{MGprocessUED} were simulated, so that
the subdominant contributions with off-shell $\tilde{g}/G^1$ are included.
We imposed the following set of cuts on the generated events:
\beq
\eta_i\,\leq\,4.0;~~\Delta R(i,j)\geq 0.4;~~p_{T,i}\,\geq\,100{~\rm GeV};
~~\met \geq 100~{\rm GeV},
\eeq{cuts}
where $i=1\ldots 4$, $j=i+1\ldots 4$ label the four (anti)quarks in each
event. The first three cuts are standard for all LHC analyses,
reflecting the finite detector coverage, separation required to define jets,
and the need to suppress the large QCD background of soft jets. The $\met$
cut is common to all searches for models where new physics events are
characterized by large missing transverse energy, such
as the MSSM and UED models under consideration. Detailed studies have shown
that this cut is quite effective in suppressing the SM backgrounds, including
both the physical background, $4j+Z$, $Z\to \nu\bar{\nu}$, and a variety of
instrumental backgrounds (see, for example, the CMS study~\cite{SUSY_CMS}).
While we have not performed an independent analysis of the SM backgrounds,
based on previous work we expect that, with a sufficiently restrictive
$\met$ cut, one will be able to obtain a large sample of new physics
events with no significant SM contamination.

\begin{figure}
\centering
\hspace{-1cm}
\begin{minipage}{8cm}
\includegraphics[width=8cm,keepaspectratio=true]{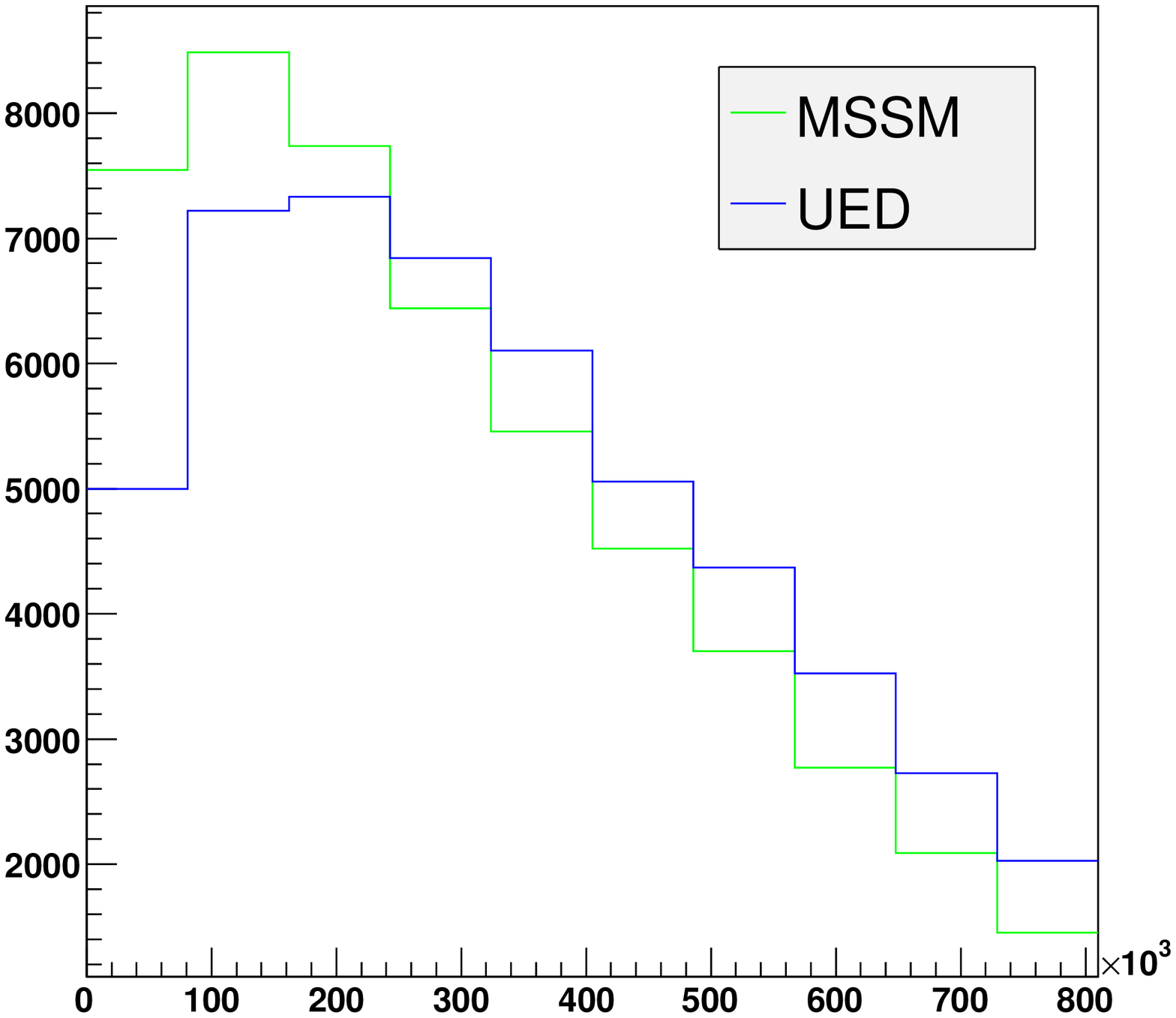}
\end{minipage}
\put(-60,-105){$s$(GeV$^2$)}
\put(-220,90){$N$}
\hspace{0.5cm}
\begin{minipage}{7.0cm}
\includegraphics[width=7.0cm,height=6.4cm]{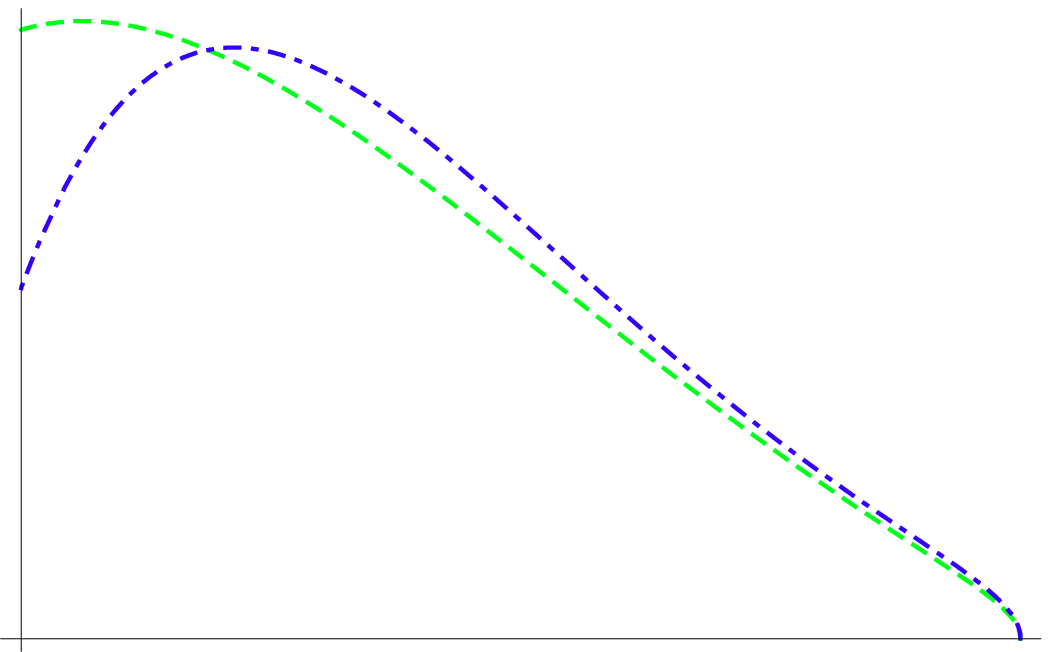}
\end{minipage}
\put(-10,-105){$s$}
\put(-215,90){$\frac{d\Gamma}{ds}$}
\caption{Left panel: Dijet invariant mass distributions from the MSSM
reaction~$pp \to qq\bar{q}\bar{q} \chi^0_1 \chi^0_1$ (green/light-gray), and
its UED counterpart~$pp \to qq\bar{q}\bar{q} B^1 B^1$ (blue/dark-gray),
including realistic experimental cuts and the combinatoric background (Monte
Carlo simulation). Right panel: Theoretical dijet invariant mass distributions
>from a single gluino/KK gluon decay with the same model parameters and no
experimental cuts.
\label{fig:MG}}
\end{figure}

The dijet invariant mass distruibutions obtained from the MSSM and UED
MC samples are shown in Fig.~\ref{fig:MG}. The distributions are
normalized to have the same total number of events, since the overall
normalization is subject to large systematic uncertainties and we do not
use any normalization information in our study. Note that for each MC event,
we include all $6$ possible jet pairings; 4 out of these correspond to
combining jets that do {\it not} come from the same decay, and thus do not
follow the theoretical distributions computed above. In Fig.~\ref{fig:MG},
we selected\footnote{This selection can be implemented in a realistic
experimental situation because $m_A-m_B$ can be measured independently.} the
jet pairs with $s\leq (m_A-m_B)^2$; all pairs with larger values of $s$ arise
>from the wrong jet pairings. However, some of the wrong jet pairs do
have $s$ in the selected range, forming a
combinatoric background to the distribution we want to study. Nevertheless,
it is clear from Fig.~\ref{fig:MG} that even after realistic cuts~\leqn{cuts}
and the combinatoric background are included, the distributions in the two
models retain their essential shape difference expected from the simplified
theoretical analysis of the previous section. Assuming that the
experimental data is described by the MSSM histogram and ignoring
systematic uncertainties, we find (using the standard $\chi^2$ test) that
about 750 events would be required to rule out the UED curve at the 99.9\%
c.l. Note that this number is smaller than those obtained in the previous
section, indicating that the performed cuts actually enhance the difference
between the MSSM and UED distributions. On the other hand, the actual
discriminating power of the analysis is likely to be somewhat lower than
this estimate, since the systematic uncertainty in the cut efficiencies was
not taken into account here.

Our parton-level analysis does not explicitly take into
account the smearing effect due to the finite jet energy and direction
resolution of a real detector. The hadronic calorimeter energy resolution
for a jet of energy E can be approximated by
\beq
\frac{\delta E}{E} \approx 0.05 \,+\, \left(\frac{1~{\rm GeV}}{E}\right)^{0.5}
\,,
\eeq{hcal}
and is in the $5-15$\% range for the jets that pass the cuts~\leqn{cuts}.
We can crudely estimate $\delta s/s$ to be of order $2\delta E/E$, evaluated
at $E=\sqrt{s}$. The fractional uncertainty of the measurement of $s$ in our
analysis is then roughly between 10\% (for points with $s\sim \smax$) and
30\% (for points with low $s$). The bin size used in Fig.~\ref{fig:MG} is
of the order of this uncertainty for large $s$, and larger for small $s$, so
we expect that the smearing introduced by binning in our analysis
provides a reasonable, if crude, description of the expected smearing
due to finite jet energy resolution. A more detailed investigation of this
effect, and other potential detector effects, would be required to fully
understand the applicability of the proposed method in a realistic
experimental situation.

%\section{From Model Discrimination to Spin Determination}

\section{Conclusions}
\setcounter{equation}{0}

In this paper, we have investigated how the dijet invariant mass distributions
>from three-body decays of a color-octed TeV-scale new particle, such as the
gluino of the MSSM and the KK gluon of the UED model, can be used to
determine the nature of this particle. The production cross section for the
color-octet state at the LHC is expected to be large, and the branching ratio
for the three-body decays is significant whenever all squarks/KK quarks
are heavier than the gluino/KK gluon. If this is the case, the dijet
invariant mass distribution can be determined accurately at the LHC.
The main complication of the analysis is that the distributions in the two
models we considered depend on a number of parameters in addition to the spin
of the decaying particle. However, even allowing for complete ignorance of
these parameters, we found the dijet invariant mass to be a very promising
tool for model discrimination.

The simplified analysis of this paper did not take into account a number of
potentially important effects. Since the particles involved are colored, the
QCD loop corrections to the decay amplitudes are expected to be significant,
and may modify the tree-level distributions we studied. Also, our analysis is
performed at the parton level and does not include detector effects. While we
expect that many systematic effects would cancel out since the analysis relies
only on the shapes of the distributions and is insensitive to the overall
normalization, a better understanding of the systematics is required. We
believe that the promising conclusions of this preliminary analysis motivate
a more detailed study of these issues.

\vskip1.0cm

\section*{Acknowledgments}

We are grateful to Matt Reece and Itay Yavin for useful discussions. This
research is supported by the NSF grant PHY-0355005. C.C. is also supported in
part by the DOE OJI grant DE-FG02-01ER41206.

%%%%%%%%%%%%%%%%%%%%%%%%%%%%%%%%%%%%%%%%%%%%%%%%%%%%%%%%%%%%%%%%%%%%%%%%%%%%%%%%%%%%%%%%%%%
\begin{appendix}

\section{Polarization Analysis of the UED case \label{app:pol}}

\begin{figure}
\centering \vspace{-2.0cm} \hspace{0.5cm}
\includegraphics[width=14.0cm]{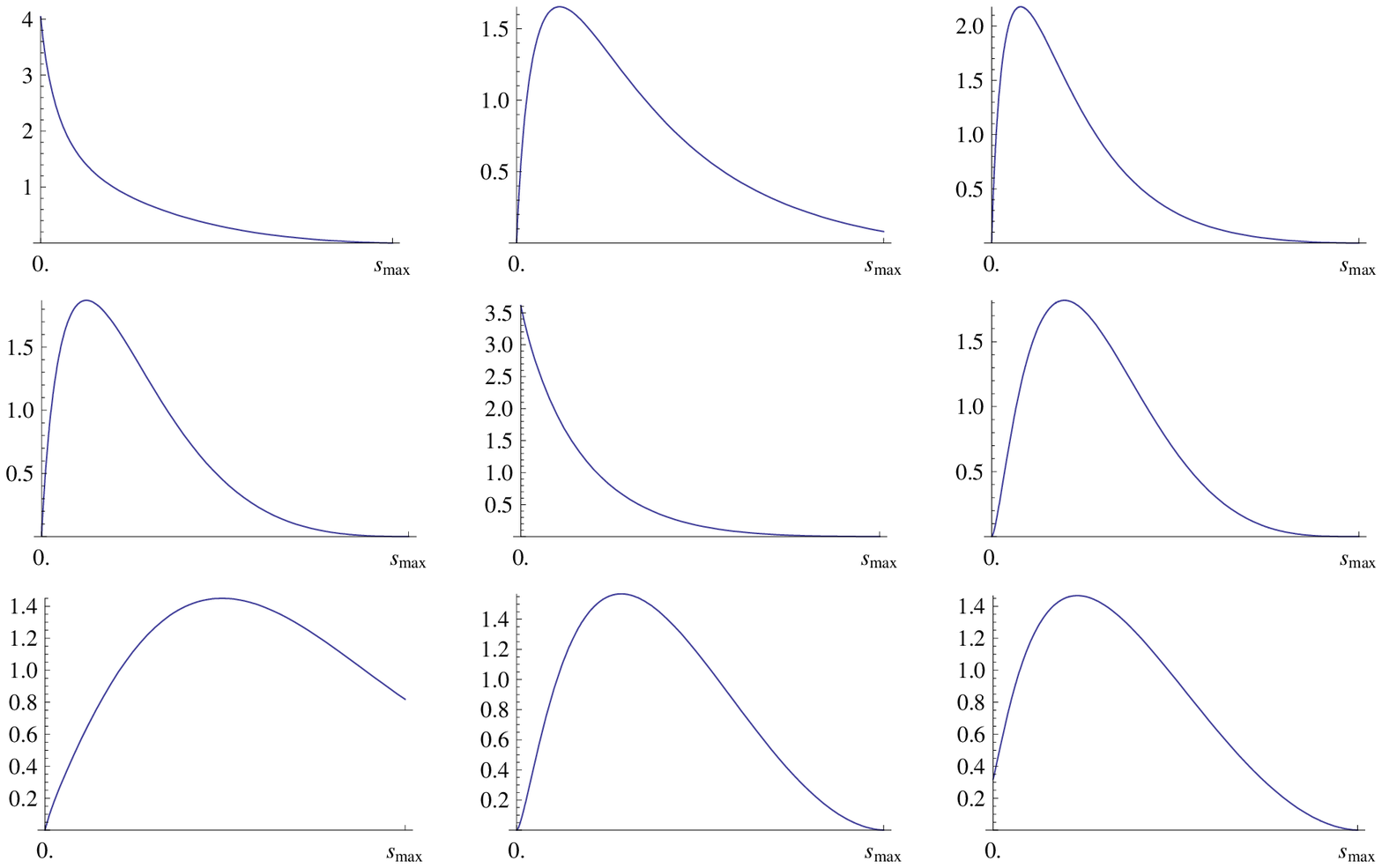}
\vspace{-2.8cm} \put(-420,290){$\epsilon^A_L$}
\put(-420,205){$\epsilon^A_-$} \put(-420,120){$\epsilon^A_+$}
\put(-340,330){$\epsilon^B_L$} \put(-200,330){$\epsilon^B_-$}
\put(-70,330){$\epsilon^B_+$} \caption{The invariant mass
distributions for the decay  of individual polarizations, divided by
the phase space distribution, for $m_B/m_A=0.1$ and $M/m_A=1.5$ in
arbitrary units. The polarization vectors are along the momentum $p_1$ of
the outgoing quark $q$. Notice that at $s=0$ only the diagonal
elements are unsuppressed due to angular momentum conservation,
resulting in a dip of the distribution. \label{fig:pol}}
\end{figure}

The main feature of the invariant mass distribution of the UED case, which 
makes it distingishable from SUSY, is the dip at $s=0$. This feature can be 
understood by analyzing the decay amplitudes of the
individual polarization components of the mother and daughter
particles and considering conservation of angular momentum. As shown
with the two toy models in Section~\ref{sec:toy}, conservation of
angular momentum can lead to suppression of the invariant mass 
distributions with respect to the pure phase space
distribution~\leqn{phasespace} at $s=0$, as well as at $s=\smax$.
The couplings in the UED case have the same chiral structure as the 
second toy model of Section~\ref{sec:toy}, with the quark and antiquark having
opposite helicities. The added complication in the UED case is that
the mother and daughter particles are massive spin one particles. We use 
$m_z(A)$ and $m_z(B)$ to denote the projections of the $A$ and $B$ spins
on the direction of the momentum $p_1$ of the quark $q$. These operators 
have eigenvalues $m_z(A), m_z(B) = -1, 0, +1$; the corresponding eigenstates 
have polarization vectors $\epsilon_-$, $\epsilon_L$, and $\epsilon_+$.
The transitions among these eigestates are described by a $3\times 3$ 
matrix of decay amplitudes. Using the UED lagrangian~\leqn{UED_fr}, we have 
evaluated these amplitudes and obtained the dijet invariant mass distribution 
corresponding to each entry.\footnote{For clarity, we only included the 
contribution of the diagrams with $Q_L^1$ in the intermediate state. 
The diagrams with $Q_R^1$ lead to distributions that are identical, up to a
parity reflection, to the ones presented here.} These distributions, 
divided out by the pure phase space distribution~\leqn{phasespace}, are 
plotted in Fig.~\ref{fig:pol}. At $s=0$ the spin projections of the 
quark-antiquark pair sums up to zero, and the final state has no angular 
momentum (see the right panel of Fig.~\ref{fig:AM}). Therefore the 
polarizations of $A$ and $B$ must be the same. This will result in a 
suppression of all 
non-diagonal components in the transition matrix at $s=0$, resulting
in a dip there. At $s=\smax$, however, the spin projections of the 
quark-antiquark pair add up to $m_z=+1$ (see Fig.~\ref{fig:AM}). Thus the 
only allowed decays at $s_\text{max}$ are the longitudinal component of 
$A$ to $m_z(B)=-1$ and $m_z(A)=+1$ to the longitudinal component of $B$. 
Both features at the ends of the distribution can be nicely observed 
in Fig.~\ref{fig:pol}.

\if
The main feature of the invariant mass distribution of the UED case
(the dip at $s=0$) can be understood by looking at the decay of the
individual polarization components of the mother and daughter
particles and considering conservation of angular momentum. As shown
with our toy models in section~\ref{sec:toy}, conservation of
angular momentum can in some kinematic cases lead to the suppression
of the invariant mass distributions compared to the pure phase space
distribution (by pure phase space we mean the invariant mass
distribution with the matrix element set one, thus considering only
pure kinematic effects and the propagator of the intermediate
particle). Figure~\ref{fig:pol} shows the invariant mass
distributions of these individual decay modes divided by the phase
space distribution.

The UED case has the same chirality assignments as our second toy
model in Fig.~\ref{fig:AM}, with the obvious added complication that
the mother and daughter particles are massive spin one particles.
Apply now the argument of Fig.~\ref{fig:AM} along the direction of
the quark $q_1$. At $s=0$ the spin projection of the quark-antiquark
pair sums up to zero, therefore the polarization of $B$ must be the
same as the one of $A$. This will result in a suppression (compared
to pure phase space) of all other components at $s=0$. At
$s=s_\text{max}$ however, the quark-antiquark pair has spin
projection $m_z=-1$. Thus the only allowed decays at $s_\text{max}$
are the longitudinal component of $A$ to $m_z(B)=+1$ and $m_z(A)=-1$
to the longitudinal polarization of $B$. Both features at the ends
of the distribution can be nicely observed in~\ref{fig:pol}.
\fi

%%%%%%%%%%%%%%%%%%%%%%%%%%%%%%%%%%%%%%%%%%%%
\section{The Kullback-Leibler distance
\label{app:KL}}
\setcounter{equation}{0}

A convenient measure to quantify how much two continuous probability
distributions differ from each other is the \emph{Kullback-Leibler distance}.
(For a recent application in the collider phenomenology context, see
Ref.~\cite{Athanasiou:2006ef}.) In this appendix, we will briefly review
this measure.

Suppose that the data sample consists of $N$ events distributed according to
the theoretical prediction of model $T$. Consider a second model, $S$, which
predicts a distribution different from $T$. We can quantify the discriminating
power of our data sample by the ratio of conditional probabilities for $S$ and
$T$ to be true, given the data:
\beq
\kappa \,=\, \frac{p(S \text{ is true}|N \text{ events from }T)}{p(T
\text{ is true}|N \text{ events from }T)}.
\eeq{kappa}
This equation can be rewritten using Bayes' theorem:
\begin{equation} \begin{split}
  \kappa &= \frac{p(S|N \text{ events from }T)}{p(T|N
\text{ events from }T)} \\
 &= \frac{p(S) p(N \text{ events from }T|S)}{p(T) p(N \text{ events from }T|T)}
\end{split} \end{equation}
where $p(S)$ and $p(T)$ are the priors -- probabilities for $S$ and $T$ to be
true before the experiment at hand is conducted. (In this paper, we assumed
that the MSSM and UED are {\it a~priori} equally likely, so we set $p(S)=p(T)=
1$.) Suppose that each event $i$ ($i=1\ldots N$) is characterized by a single
variable $s_i$ (in our case, the dijet invariant mass). Since the $N$ events
are independent, we have
\begin{equation} \begin{split}
\kappa &= \frac{p(S)}{p(T)} \frac{\prod_{i=1}^N p(s_i^{(T)}|S)}{\prod_{i=1}^N
p(s_i^{(T)}|T)}\\
      &= \frac{p(S)}{p(T)} \exp \left( \sum_{i=1}^N \log \left(
\frac{p(s_i^{(T)}|S)}{p(s_i^{(T)}|T)} \right) \right).
\end{split} \end{equation}
For large $N$ we can approximate $\sum_N \approx \int ds \frac{dN}{ds}$ and
use the normalization condition $\frac{dN}{ds}= N p(s|T)$ to obtain
\begin{equation} \begin{split}
\kappa &\approx  \frac{p(S)}{p(T)} \exp \left( N \int \hspace{-0.1cm}ds \log
\left( \frac{p(s|S)}{p(s|T)} \right) p(s|T) \right) \\
       &= \frac{p(S)}{p(T)} \exp \left( - N \hspace{0.1cm}\text{KL}(T,S)
\right),
\end{split} \end{equation}
where the \emph{Kullback-Leibler distance} (also called \emph{relative
entropy}) is defined as
\begin{equation}
    \text{KL}(T,S):=\int \hspace{-0.1cm}ds \log \left( \frac{p(s|T)}{p(s|S)}
\right) p(s|T).
\end{equation}
It follows that the number of events needed to constrain the probability of
model $S$ being true, relative to the probability of $T$ being true, to be
less than $1/R$, is given by
\begin{equation}
  N \approx \frac{\log R + \log \frac{p(S)}{p(T)}}{\text{KL}(T,S)}.
\end{equation}
This number provides a convenient and physically meaningful measure of how
different the $S$ and $T$ distributions are.

Two properties of the Kullback-Leibler distance are worth mentioning in our 
context.
First, while this is not manifest from its definition, the KL distance is
non-negative, and zero if and only if $T=S$. Second, it is invariant under
transformations $s \rightarrow f(s)$: for example, it does not matter whether
we consider the jet invariant mass distribution in terms of $s$ or
$m_{jj}=\sqrt{s}$.

\end{appendix}

\end{document}